\documentclass[a4paper]{article}

\usepackage{INTERSPEECH2022}

\usepackage{xcolor,soul,framed} 
\usepackage{amsmath,bm}
\usepackage{url}            

\title{Perceptual Contrast Stretching on Target Feature for Speech Enhancement}
\name{Rong Chao$^{1,2}$, Cheng Yu$^2$, Szu-Wei Fu$^3$, Xugang Lu$^4$, Yu Tsao$^2$}
\address{
  $^1$CSIE, NCKU, Taiwan \
  $^2$CITI, Academia Sinica, Taiwan \
  $^3$Microsoft Corporation \
  $^4$NICT, Japan
  }
\email{f14071075@gs.ncku.edu.tw, yu.tsao@citi.sinica.edu.tw}

\begin{document}

\maketitle

\begin{abstract}

Speech enhancement (SE) performance has improved considerably owing to the use of deep learning models as a base function. Herein, we propose a perceptual contrast stretching (PCS) approach to further improve SE performance. The PCS is derived based on the critical band importance function and is applied to modify the targets of the SE model. Specifically, the contrast of target features is stretched based on perceptual importance, thereby improving the overall SE performance. Compared with post-processing-based implementations, incorporating PCS into the training phase preserves performance and reduces online computation. Notably, PCS can be combined with different SE model architectures and training criteria. Furthermore, PCS does not affect the causality or convergence of SE model training. Experimental results on the VoiceBank-DEMAND dataset show that the proposed method can achieve state-of-the-art performance on both causal (PESQ score = 3.07) and noncausal (PESQ score = 3.35) SE tasks.

\end{abstract}
\noindent\textbf{Index Terms}: Speech enhancement, contrast stretching, perceptual importance

\section{Introduction}

Speech enhancement (SE) is performed to remove noise components from noisy speech to improve speech quality and intelligibility. SE has been used at an important front-end of many speech-related studies, such as automatic speech recognition \cite{weninger2015speech, zhang2017speech}, speaker recognition \cite{michelsanti2017conditional, taherian2020robust}, and assistive listening devices \cite{wang2017deep, lesica2021harnessing}. Traditionally, SE algorithms are typically designed based on the assumptions of speech and noise signals. Notable approaches include those presented in \cite{boll1979suppression} and \cite{priori1}. These approaches are effective in some stationary noise scenarios, wherein the signals conform to the assumptions introduced. However, in most real-world noisy scenarios, time-varying noise exhibits nonstationary properties, resulting in the suboptimal performance of these conventional SE methods. 

In recent years, deep learning (DL) has been widely used in various research fields, including SE.
Using DL models as a base mapping function has notably improved SE \cite{lu2013speech, xu2014regression,  liu2014experiments, han2015learning}. Although these DL-based methods achieve satisfactory performance under the testing conditions associated with the training data, their performance degrades when they are operated under unexpected conditions, attributed to two factors: first, the network architecture may not adequately consider the sequential nature of the speech signals; second, a regression function optimized by L1/L2 distance-based objective functions may average out important signal patterns, which can result in low precision of enhanced speech.

Numerous approaches have been proposed to further improve DL-based SE systems, including those that aims to determine suitable acoustic features to improve SE, such as waveforms \cite{fu2018end} and complex spectral features \cite{williamson2015complex}. Additionally, advanced networks have been proposed to model sequential signals more accurately, such as recurrent neural networks \cite{huang2015joint}, fully convolutional networks \cite{fu2018end}, long-short term memory \cite{LSTM2}, transformers \cite{kim2020t}, and generative adversarial networks \cite{pascual2017segan, fu2019metricgan}. In another approach, advanced objective functions are derived to provide more accurate training to achieve the desired speech quality or intelligibility. Notable examples include differentiable speech metrics \cite{fu2019metricgan, fu2018end}, and deep feature losses \cite{germain2019speech, hsieh2020improving}, where losses are computed in representative or discriminative feature spaces. Aside from the above-mentioned approaches, this study investigates another direction to improve the SE performance: modifying the target features of the SE model and post-processing (PP).

PP has been derived to further modify enhanced speech to match the statistical properties of clean speech \cite{xu2014regression, valin2020perceptually, chen2020truth}. Experimental results show that such an approach can further sharpen the structure of enhanced speech and suppress residual noise. Moreover, PP is compatible with any system for further improving SE. Accordingly, we propose perceptual contrast stretching (PCS)—a novel method to enhance the contrasts of the target features of an SE model. PCS can be implemented as PP for SE or incorporated into the SE training phase, where the implementation of the latter can avoid an increase computation cost during inference.



Gamma-correction approaches have proven to be effective for image enhancement \cite{rahman2016adaptive}. Based on these approaches, the characteristics of the human auditory system are employed in PCS. In this study, we first examined the effectiveness of PCS by incorporating it into the SE training phase. Next, we implemented it as a PP, and notable improvements were achieved. Specifically, PCS stretches the contrasts of the target features in the training data based on a set of auditory weights. The weights are designed based on the critical band importance \cite{pavlovic2018sii}, which is perceptually correlated with the human auditory system. The proposed PCS offers three major advantages: First, it is compatible with different SE systems (conventional or DL based). Second, it does not require additional parameters in the SE model. Third, it does not affect the causal property of the causal SE models.
\begin{figure*}
 \centering
 \includegraphics[width=\linewidth]{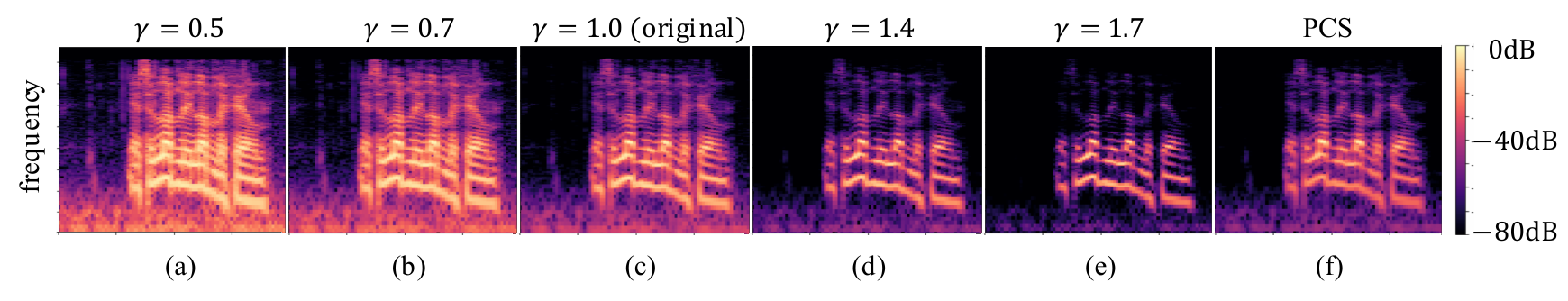}
 \caption{Normalized time-frequency feature (spectral magnitude) is stretched by different gamma values. The original clean feature ($\gamma$ = 1) is shown in (c); stretching based on different $\gamma$ values is shown in (a) to (e); (f) shows the proposed PCS.} 
 \label{fig:gammagraph}
\end{figure*}
We compared PCS with several different contrast-stretching strategies in our experiments. The evaluation scores show that PCS outperformed the other feature enhancement approaches. State-of-the-art (SOTA) results (PESQ score = 3.35) were achieved when the best SE model was used.

\section{Related Studies}
In this section, we introduce two primary categories of related studies. Our proposed PCS training strategy for SE was designed based on these previous studies. 

\subsection{Gamma correction}
\label{ssec:gammacorrection}

The proposed PCS was partially inspired by image processing. First, we introduce gamma correction \cite{poynton2012digital}, widely adopted as a contrast-enhancement approach in computer vision research. Based on human vision systems, the brightness perception of the eye from graphical information is affected by the brightest region of the image. Specifically, the relationship between perceived and physical brightness can be derived as a nonlinear transfer function. The gamma correction was designed based on this function and can be derived as a power-law operation, expressed as follows:


\begin{equation}
\label{equation:gamma}
V_{out} = AV_{in}^\gamma
\end{equation}
where $V_{in}$ is the input signal value, $\gamma$ the modulation parameter, $A$ the scaling function (typically a constant), and $V_{out}$ the output signal value. For example, the standard red-green-blue (sRGB), widely used in monitors and printers, uses a gamma value of 2.2 ($\gamma = 2.2$) in its transfer function to provide better perception. Another color space, i.e., the Digital Cinema Initiatives - Protocol 3 (DCI-P3), uses a gamma value of 2.6 ($\gamma = 2.6$). In these operations, the input signals are normalized to a range between 0 and 1 to ensure that the boundary values and the minimal and maximal values of the input signal remain invariant after the operation. In recent SE research, Zhang et al. \cite{zhang2021low} applied a weighting mechanism to the training target, similar to gamma correction using a dynamic scaling function (i.e., based on the ratio of input and target features).  



\subsection{Critical band importance}
\label{ssec:criticalband}
In human auditory systems, the importance of the signal components varies based on the frequency region. More specifically, humans can perceive differences in frequency bands ranging from 400 to 4400 Hz better than in other frequency bands. Consequently, a set of critical band importance weights was measured and defined. Some conventional SE approaches adopted critical bands to perform spectral subtraction \cite{singh1998speech}, whereas some DL-based SE approaches adopted critical band importance to improve their model \cite{liu2018bone}. These approaches demonstrate improvements in perceptual evaluation scores compared with the baseline approaches. In this study, we demonstrate that SE can be further improved to achieve SOTA performance by combining gamma correction with critical band importance.

\section{PCS on target feature for speech enhancement}
In this section, the derivation of the proposed PCS on the target feature for SE based on two related studies is presented.

\subsection{Auditory nonlinearity}
Similar to the human vision system, the human auditory system exhibits a nonlinear relationship with speech signals. The sound pressure level (SPL) is measured in decibels (dB) as follows:

\begin{equation}
\beta(dB) = 10log_{10}(I/I_{0})
\end{equation}

This equation standardizes the relationship between physical loudness and perceived loudness. The notations $\beta$, $I$, and $I_{0}$ denote the dB level, measured signal power, and reference signal power, respectively. The SPL of the human auditory and visual systems
exhibits a similar property because the human eye perceives brightness based on reference to the brightest region of an image, as mentioned in \ref{ssec:gammacorrection}. Thus, we designed a transfer function suitable for the human auditory systems.

\begin{figure}[h]
 \centering
 \includegraphics[width=\linewidth]{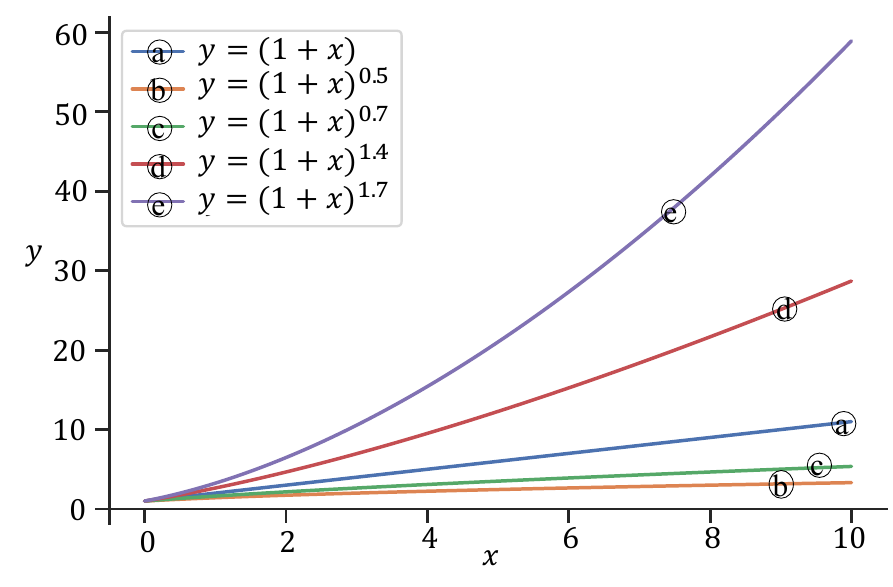}
 \caption{Scaling curves of our modified gamma correction using different gamma values, where $x$ and $y$ represent input and output, respectively.}  
 \label{fig:gammascale}
\end{figure}

\begin{figure*}[t]
 \centering
 \includegraphics[width=\linewidth]{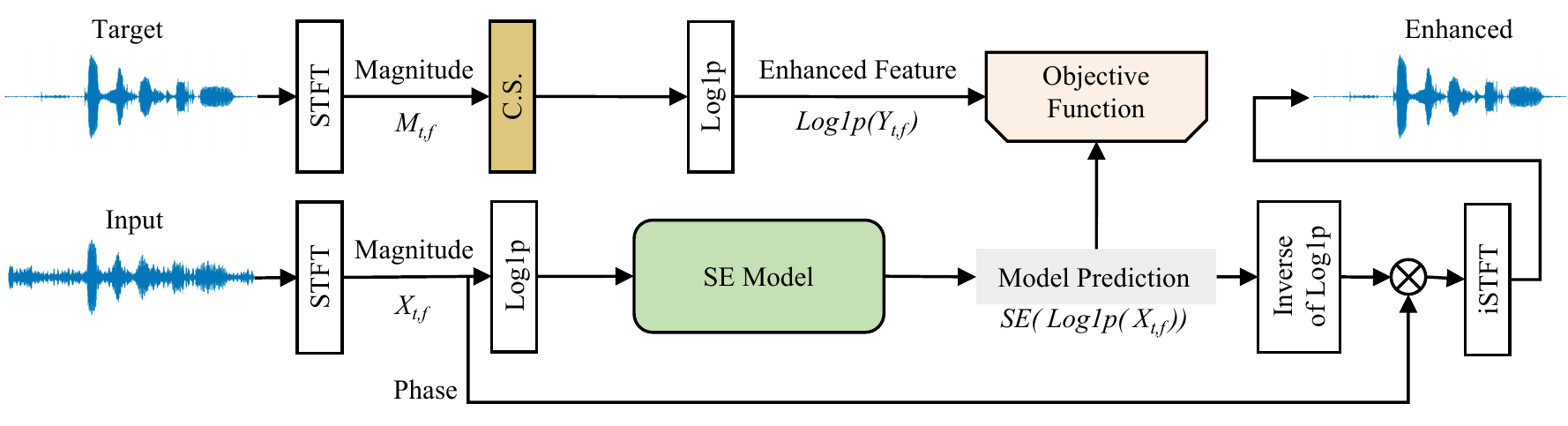}
 \caption{High-level system block diagram of proposed perceptual contrast stretching training strategy for SE, where the brown block C.S. represent contrast-stretching step.} 
 \label{fig:system}
 \vspace{-0.3cm}
\end{figure*}

\subsection{Modified gamma correction}
We begin by implementing gamma correction on time-frequency features (spectrogram) of speech signals to enhance the contrast.
Based on Eq. \ref{equation:gamma}, the range of input features must be suitable for our task. In Eq. \ref{equation:gamma}, the range of $V_{in}$ is regulated between $[0, 1]$. However, the range of speech features cannot be regulated within this range. Hence, a new regulation is required, wherein the input values range between $[0, V]$ (here, $V$ represents the maximum value of the input features). The processed signal was normalized when it was recovered in the waveform domain. Subsequently, the gamma correction equation for the time-frequency feature of speech signals was modified as follows:
\begin{equation}
\label{equation:tfgamma}
Y_{t,f} = A(M_{t,f})^\gamma
\end{equation}
where the value of input feature $M_{t,f}$ ranges from $[0, M]$. The notations $Y_{t,f}$, $A$, $\gamma$, and $M_{t,f}$ denote the modified feature, scaling function, gamma value, and input feature, respectively.


Furthermore, the training features of our SE models were moved to the $log(1 + p)$ domain ($Log1p$ features) \cite{fu2020boosting}. We can thus derive Eq. \ref{equation:tfgamma} as follows:
\begin{equation}
\label{equation:log1pgamma}
\begin{aligned}
    log1p(Y_{t,f}) = \log (1+Y_{t,f}) = \gamma * \log (1+M_{t,f})
\end{aligned} 
\end{equation}
where the scaling function $A$ is $(1+1/M_{t,f})^{\gamma} - (1/M_{t,f})^{\gamma}$ in this case, which is dynamic and based on $M_{t,f}$. As shown in Fig \ref{fig:gammascale}, the modified gamma correction attenuates the features (decreases the contrast) when $\gamma < 1$, and increases the value of the features (enhances the contrast) when $\gamma > 1$. This effect is similarly observed on the speech spectrogram shown in Fig. \ref{fig:gammagraph}, where (a) and (b) contain more blurry patterns, whereas (d) and (e) contain more contrastive patterns. This also suggests that feature enhancement can be implemented using different $\gamma$ values to serve specific purposes. 

\subsection{PCS}
Contrast-stretching was applied to perceptually enhance the target features on the training data. The waveform was first processed via short-time Fourier transform (STFT), wherein the phase component was excluded from the input stream for inverse STFT (iSTFT) (see Fig. \ref{fig:system}). Subsequently, we applied contrast stretching to the target stream feature for the enhanced feature $Y_{t,f}$. The $Log1p$ feature was then obtained using the feature (here, $Y_{t,f}$ and $X_{t,f}$ denote the target and input streams, respectively). Consequently, the loss $L$ can be computed as:
\begin{equation}
\begin{aligned}
    L = D(SE(Log1p(X_{t,f})), \ Log1p(Y_{t,f}))
\end{aligned}
\end{equation}
where $D(\cdot)$ denotes the objective functions, and $SE(\cdot)$ denotes the transformation by the SE. To determine the best performance afforded using a fixed $\gamma$, we tested several different hyperparameters ($\gamma = 0.5$–$2.0$, with a step size of $0.1$) in our validation set and evaluated their effectiveness. Experimental results on the VoiceBank-DEMAND dataset show that the best perceptual scores were achieved using $\gamma = 1.4$. Hence, we adopted $1.4$ as the fundamental $\gamma$ value. 
\begin{table}[h]
\centering
\caption{Critical band importance and proposed PCS.}
\label{tab:table_PCS}
\begin{tabular}[\linewidth]{|c|c|c|}
\hline
Frequency bands (Hz) & BIF & \boldsymbol{$\gamma_{PCS}$} \\
\hline 
0$\sim$100           & 0.000 & 1.0000  \\
100$\sim$200         & 0.010 & 1.0702  \\
200$\sim$300         & 0.026 & 1.1825  \\
300$\sim$400         & 0.041 & 1.2877   \\
400$\sim$4400        & 0.057 & 1.4    \\
4400$\sim$5300       & 0.046 & 1.3228   \\
5300$\sim$6400       & 0.034 & 1.2386   \\
6400$\sim$7700       & 0.023 & 1.1614   \\
7700$\sim$9500       & 0.011 & 1.0772  \\
\hline
\end{tabular}
\end{table}

We reviewed Cochlea's knowledge in Sec. \ref{ssec:criticalband}, its applications, and its importance in perceptual scores when applied to SE tasks. To further improve our contrast stretching for better perceptual performance, we designed feature enhancement based on the critical band importance. As most SE approaches adopt time–frequency features, we designed a feature enhancement that weights the features based on their frequencies. The band importance function (BIF) \cite{ansi1997s3} is listed in Table \ref{tab:table_PCS}. We designed the PCS based on the BIF and rescaled it into the range [1.0, 1.4]. The rescaling function is formulated as follows:
\begin{equation}
\begin{aligned}
    \gamma_{PCS}[k] =  \frac{( \gamma - PCS_{min} )}{( BIF_{Max} - BIF_{min})}  * BIF[k] + PCS_{min} 
\end{aligned}
\end{equation}
where $k$ denotes the index of the frequency bands, $\gamma_{PCS}[k]$ and $BIF[k]$ are the $\gamma$ values of PCS and BIF at band $k$, respectively. $BIF_{Max}$ and $BIF_{min}$ denote the maximum and minimum values of $BIF$, respectively. Meanwhile, $PCS_{min}$ was set to $1$, and $\gamma = 1.4$ was adopted. As shown in Fig. \ref{fig:gammagraph}(f), the proposed PCS\footnote{The actual frequency band regions we used to implement PCS are slightly different from Table \ref{tab:table_PCS} owing to the STFT limitations. Detailed settings can be fount at \url{https://github.com/RoyChao19477/PCS/PCS}.} is effective. We can sharpen the formant peaks by applying PCS, where Fig. \ref{fig:gammagraph}(f) is more contrastive compared with those presented in Figs. \ref{fig:gammagraph}(a) and \ref{fig:gammagraph}(b), but does not indicate severe distortions as in Figs. \ref{fig:gammagraph}(d) and \ref{fig:gammagraph}(e).

\section{Experiments}
\label{sec:exp}

\subsection{Experimental setup}
\label{ssec:network}

We evaluated our method using well-known network architectures, including the transformer \cite{fu2020boosting}, CRNN \cite{zhao2018convolutional}, MetricGAN+ \cite{fu2021metricgan+} (from SpeechBrain \cite{ravanelli2021speechbrain}), and DPT-FSNet \cite{dang2021dpt}. The transformer contains four convolutional encoder layers, eight self-attention heads, and a fully connected decoder layer. The CRNN comprises CNN layers, with one bidirectional long short-term memory (BLSTM) layer and two fully connected layers. The MetricGAN+ comprises a BLSTM-based generator with two bidirectional LSTM layers and a CNN-based discriminator. The dual-path transformer-based full-band and sub-band fusion network (DPT-FSNet) is a dual-path architecture that uses an improved transformer.

To compare the proposed PCS with other methods, we used a publicly available dataset, VoiceBank-DEMAND, to evaluate the SE. The Voice Bank-DEMAND dataset is widely used as a benchmark for monaural SE approaches. The training set with 11572 utterances comprises 28 speakers corrupted with four signal-to-noise ratio (SNR) levels (15, 10, 5, and 0 dB). The test sets set with 824 utterances comprises two speakers corrupted at four SNR levels (17.5, 12.5, 7.5, and 2.5 dB). Details regarding this dataset are available in \cite{valentini2017noisy}. 

\subsection {Comparison of different feature enhancement methods}
\label{ssec:comparison}

As contrast-stretching for image processing can be considered a feature enhancement method for SE, we compared our proposed PCS with a fixed gamma of 1.4 with three typically used contrast enhancement methods—min-max normalization, histogram equalization (HE) \cite{pizer1987adaptive}, and adaptive equalization (AE) \cite{qureshi1985adaptive}. These methods were adopted in the spectrum domain as contrast-stretching methods. HE was implemented along the time axis of each frequency band. A causal transformer SE model was used to evaluate the effectiveness of these methods. 
Feature enhancement was applied to the spectral domain and then transferred to the log1p features space in these experiments.

\begin{table}
\caption{SE models with different feature enhancement methods on VoiceBank-DEMAND.}
\vspace{-0.2cm}
\label{table_1}
\centering
\begin{tabular*}{\linewidth}{l||lllll} 
\hline
                  & PESQ  & STOI  & CSIG  & COVL  \\ 
\hline
Noisy             & 1.97 & 0.92 & 3.34 & 2.63 \\
\hline
Wiener \cite{pascual2017segan} & 2.22 & - & 3.23 & 2.67 \\
Conv-TasNet \cite{koyama2020exploring} & 2.53 & - & 3.95 & 3.23 \\ 
Demucs \cite{defossez2020real} & 2.93  & \textbf{0.95}   & 4.22  & 3.52 \\
T \cite{fu2020boosting} & 2.76 & 0.94 & 4.10 & 3.44 \\
T + min-max       & 2.80 & 0.93 & 4.09 & 3.45 \\
T + HE  & 2.20 & 0.93 & 3.04 & 2.60 & \\
T + AE            & 2.82 & 0.94 & 4.12 & 3.48 \\
\hline
\hline
\textbf{T + $\gamma$=1.4}   & 2.90 & 0.94 & 4.18 & 3.55 \\
\textbf{T + \boldsymbol{$\gamma$}=PCS} & \textbf{3.07} & 0.94 & \textbf{4.26} & \textbf{3.67} \\
\hline
\end{tabular*}
\end{table}
Table \ref{table_1} shows that the causal transformer (T) can be improved notably via PCS method, with a 0.31 PESQ score improvement. Moreover, except for HE, which failed to converge in training, the enhancement results are beneficial when feature enhancement approaches are used. To evaluate the effectiveness of the causal transformer with PCS, we tested performance of several causal models (e.g., causal DEMUCS \cite{defossez2020real} and Conv-TasNet (scores from \cite{koyama2020exploring})).
To the best of our knowledge, the proposed method outperformed other causal SE methods on this dataset in terms of the PESQ, CSIG, and COVL scores.

\begin{table}[h]
\caption{Different models with PCS on VoiceBank-DEMAND.}
\label{table_2}
\centering
\begin{tabular}{l||lllll} 
\hline
                              & PESQ           & STOI  & CSIG    &  COVL  & Cau.  \\
\hline
Noisy             & 1.97 & 0.92 & 3.34 & 2.63 & - \\
\hline
SEGAN \cite{pascual2017segan} & 2.16 & - & 3.48 & 2.80 & No \\
T (c) \cite{fu2020boosting} & 2.76          & 0.94 & 4.10 & 3.44 & Yes     \\
T (c) \textbf{+ PCS}    & \textbf{3.07} & 0.94 & \textbf{4.26} & \textbf{3.67} & \textbf{Yes}     \\
T (nc) \cite{fu2020boosting} & 2.84          & 0.94 & 4.20 & 3.51 & No      \\
T (nc) \textbf{+ PCS}    & \textbf{3.15} & 0.94 & \textbf{4.34} & \textbf{3.75} & No      \\
CRNN \cite{zhao2018convolutional} & 2.83          & 0.94 & 4.18 & 3.51 & No      \\
CRNN \textbf{+ PCS}     & \textbf{3.11} & 0.94 & \textbf{4.31} & \textbf{3.72} & No      \\
MGAN+ \cite{fu2021metricgan+}    & 3.15 & 0.93 & 4.14  & 3.64  & No      \\ 
MGAN+ \textbf{+ PCS}    & \textbf{3.21} & 0.93 & \textbf{4.15} & \textbf{3.67} & No      \\
DPT \cite{dang2021dpt}  & 3.33 & 0.96 & \textbf{4.58} & \textbf{4.00} & No      \\
DPT*                    & {3.11} & 0.95 & 4.30 & 3.72 & No      \\
DPT* \textbf{+ PCS}     & \textbf{3.35} & 0.95 & \textbf{4.43} & \textbf{3.92} & No      \\
\hline
\end{tabular}
\end{table}

\subsection {Effectiveness using different SE models}

Table \ref{table_2} shows that the performance of different SE models, including the causal transformer (T(c)), non-causal transformer (T(nc)), CRNN, MetricGAN+ (MGAN+), and DPT-FSNet\footnote{The contrast-stretched target features are transferred back to the waveform domain as training targets.} (We used a frame size of 64 and a single batch size in all DPT-FSNet reproduction experiments and denote it as DPT*), can be improved by applying PCS. To the best of our knowledge, The DPT* + PCS achieved SOTA performance with a PESQ score of 3.35 and outstanding scores for other evaluation metrics. 
In general, we can infer from the improvements presented in Table \ref{table_2} that PCS is a general and effective contrast stretching training strategy for DL-based SE approaches.

\subsection {Effectiveness of PP}
PCS can be implemented as a PP module, aiming to further improve enhanced speech from an SE model. The results are presented in Table \ref{table_3}. Note that for some CS methods (e.g., HE), information from the entire spectrogram is required, causing the SE system noncausal. The operation of PCS is similar to gamma correction; therefore the causality of the SE model will not be changed when PCS is applied as PP. From Table \ref{table_3}, using PCS as a PP module can also improve the SE performance effectively. 
When comparing the results of Tables. \ref{table_2} and \ref{table_3}, using PCS as a PP and applying PCS to the target feature enhancement yield comparable improvements. Especially, implementing PCS as a PP module allows it to be used alone (Noisy\textbf{+PP-PCS}) and combined with conventional SE methods (e.g., Wiener\textbf{+PP-PCS}); the detailed results and codes can be found at \url{https://github.com/RoyChao19477/PCS}.


\begin{table}[h]
\caption{PP-PCS with different SE models on the VoiceBank-Demand dataset. "Noisy" denotes original speech without SE.}
\label{table_3}
\centering
\begin{tabular}{l||llllll} 
\hline
                  & PESQ & STOI & CSIG & COVL    \\
\hline
Noisy             & 1.97 & 0.92 & 3.34 & 2.63    \\
Noisy \textbf{+ PP-PCS} & \textbf{2.47} & 0.92 & \textbf{3.63}  & \textbf{3.03}  \\
Wiener             & 2.22 & 0.91 & 3.21  & 2.65        \\
Wiener\textbf{+ PP-PCS}      & \textbf{2.63} & 0.91 & \textbf{3.39}  & \textbf{2.95}        \\
MGAN+ \textbf{+ PP-PCS}    & 3.20 & 0.92 & 4.08 & 3.63    \\
DPT* \textbf{+ PP-PCS}     & \textbf{3.30} & \textbf{0.95} & \textbf{4.35} & \textbf{3.84}    \\
\hline
\end{tabular}
\end{table}

\section{Conclusions}
\label{sec:conclusion}
We proposed a PCS for target features to further boost SE performance. PCS exerts a perceptual emphasis on target features to overcome the average-out problem (not precise) caused by distance-based objective functions. Three major contributions of the proposed PCS are noted: First, PCS is compatible with different SE models (both conventional and DL based). Second, no additional parameters are required in the SE model. Third, it does not affect the causality of causal SE models. We conclude that the proposed PCS can further improve the performance of previous SOTA SE models with an efficient operation of the target features. 
To the best of our knowledge, the causal transformer + PCS (causal) and DPT* + PCS (noncausal) approaches achieved the best PESQ score and competitive scores in other metrics on the Voice Bank-DEMAND dataset.




\newpage

\bibliographystyle{IEEEtran}

\bibliography{mybib}

\begin{thebibliography}{10}
\providecommand{\url}[1]{#1}
\csname url@samestyle\endcsname
\providecommand{\newblock}{\relax}
\providecommand{\bibinfo}[2]{#2}
\providecommand{\BIBentrySTDinterwordspacing}{\spaceskip=0pt\relax}
\providecommand{\BIBentryALTinterwordstretchfactor}{4}
\providecommand{\BIBentryALTinterwordspacing}{\spaceskip=\fontdimen2\font plus
\BIBentryALTinterwordstretchfactor\fontdimen3\font minus
  \fontdimen4\font\relax}
\providecommand{\BIBforeignlanguage}[2]{{%
\expandafter\ifx\csname l@#1\endcsname\relax
\typeout{** WARNING: IEEEtran.bst: No hyphenation pattern has been}%
\typeout{** loaded for the language `#1'. Using the pattern for}%
\typeout{** the default language instead.}%
\else
\language=\csname l@#1\endcsname
\fi
#2}}
\providecommand{\BIBdecl}{\relax}
\BIBdecl

\bibitem{weninger2015speech}
F.~Weninger, H.~Erdogan, S.~Watanabe, E.~Vincent, J.-L. Roux, J.-R. Hershey,
  and B.~Schuller, ``Speech enhancement with {LSTM} recurrent neural networks
  and its application to noise-robust {ASR},'' in \emph{Proc. LVA/ICA}, 2015.

\bibitem{zhang2017speech}
X.~Zhang, Z.-Q. Wang, and D.~Wang, ``A speech enhancement algorithm by
  iterating single-and multi-microphone processing and its application to
  robust {ASR},'' in \emph{Proc. ICASSP}, 2017.

\bibitem{michelsanti2017conditional}
D.~Michelsanti and Z.-H. Tan, ``Conditional generative adversarial networks for
  speech enhancement and noise-robust speaker verification,'' in \emph{Proc.
  INTERSPEECH}, 2018.

\bibitem{taherian2020robust}
H.~Taherian, Z.-Q. Wang, J.~Chang, and D.~Wang, ``Robust speaker recognition
  based on single-channel and multi-channel speech enhancement,''
  \emph{IEEE/ACM Transactions on Audio, Speech, and Language Processing},
  vol.~28, pp. 1293--1302, 2020.

\bibitem{wang2017deep}
D.~Wang, ``Deep learning reinvents the hearing aid,'' \emph{IEEE spectrum},
  vol.~54, no.~3, pp. 32--37, 2017.

\bibitem{lesica2021harnessing}
N.~A. Lesica, N.~Mehta, J.~G. Manjaly, L.~Deng, B.~S. Wilson, and F.-G. Zeng,
  ``Harnessing the power of artificial intelligence to transform hearing
  healthcare and research,'' \emph{Nature Machine Intelligence}, vol.~3,
  no.~10, pp. 840--849, 2021.

\bibitem{boll1979suppression}
S.~Boll, ``Suppression of acoustic noise in speech using spectral
  subtraction,'' \emph{IEEE Transactions on acoustics, speech, and signal
  processing}, vol.~27, no.~2, pp. 113--120, 1979.

\bibitem{priori1}
P.~Scalart and J.~V. Filho, ``Speech enhancement based on a priori signal to
  noise estimation,'' in \emph{Proc. ICASSP}, 1996.

\bibitem{lu2013speech}
X.~Lu, Y.~Tsao, S.~Matsuda, and C.~Hori, ``Speech enhancement based on deep
  denoising autoencoder.'' in \emph{Proc INTERSPEECH}, 2013.

\bibitem{xu2014regression}
Y.~Xu, J.~Du, L.-R. Dai, and C.-H. Lee, ``A regression approach to speech
  enhancement based on deep neural networks,'' \emph{IEEE/ACM Transactions on
  Audio, Speech, and Language Processing}, vol.~23, no.~1, pp. 7--19, 2014.

\bibitem{liu2014experiments}
D.~Liu, P.~Smaragdis, and M.~Kim, ``Experiments on deep learning for speech
  denoising,'' in \emph{Proc. INTERSPEECH}, 2014.

\bibitem{han2015learning}
K.~Han, Y.~Wang, D.~Wang, W.~S. Woods, I.~Merks, and T.~Zhang, ``Learning
  spectral mapping for speech dereverberation and denoising,'' \emph{IEEE/ACM
  Transactions on Audio, Speech, and Language Processing}, vol.~23, no.~6, pp.
  982--992, 2015.

\bibitem{fu2018end}
S.-W. Fu, T.-W. Wang, Y.~Tsao, X.~Lu, and H.~Kawai, ``End-to-end waveform
  utterance enhancement for direct evaluation metrics optimization by fully
  convolutional neural networks,'' \emph{IEEE/ACM Transactions on Audio,
  Speech, and Language Processing}, vol.~26, no.~9, pp. 1570--1584, 2018.

\bibitem{williamson2015complex}
D.~Williamson, Y.~Wang, and D.~Wang, ``Complex ratio masking for monaural
  speech separation,'' \emph{IEEE/ACM transactions on audio, speech, and
  language processing}, vol.~24, no.~3, pp. 483--492, 2015.

\bibitem{huang2015joint}
P.-S. Huang, M.~Kim, M.~Hasegawa-Johnson, and P.~Smaragdis, ``Joint
  optimization of masks and deep recurrent neural networks for monaural source
  separation,'' \emph{IEEE/ACM Transactions on Audio, Speech, and Language
  Processing}, vol.~23, no.~12, pp. 2136--2147, 2015.

\bibitem{LSTM2}
Z.~Chen, S.~Watanabe, H.~Erdogan, and J.~R. Hershey, ``Speech enhancement and
  recognition using multi-task learning of long short-term memory recurrent
  neural networks,'' 2015.

\bibitem{kim2020t}
J.~Kim, M.~El-Khamy, and J.~Lee, ``T-gsa: {Transformer} with gaussian-weighted
  self-attention for speech enhancement,'' in \emph{Proc. ICASSP}, 2020.

\bibitem{pascual2017segan}
S.~Pascual, A.~Bonafonte, and J.~Serr{\`a}, ``Segan: Speech enhancement
  generative adversarial network,'' in \emph{Proc. INTERSPEECH}, 2017.

\bibitem{fu2019metricgan}
S.-W. Fu, C.-F. Liao, Y.~Tsao, and S.-D. Lin, ``Metricgan: Generative
  adversarial networks based black-box metric scores optimization for speech
  enhancement,'' in \emph{Proc. ICML}, 2019.

\bibitem{germain2019speech}
F.~Germain, Q.~Chen, and V.~Koltun, ``Speech denoising with deep feature
  losses,'' in \emph{Proc. INTERSPEECH}, 2019.

\bibitem{hsieh2020improving}
T.-A. Hsieh, C.~Yu, S.-W. Fu, X.~Lu, and Y.~Tsao, ``Improving perceptual
  quality by phone-fortified perceptual loss using {Wasserstein Distance} for
  speech enhancement,'' in \emph{Proc. INTERSPEECH}, 2021.

\bibitem{valin2020perceptually}
J.-M. Valin, U.~Isik, N.~Phansalkar, R.~Giri, K.~Helwani, and A.~Krishnaswamy,
  ``A perceptually-motivated approach for low-complexity, real-time enhancement
  of fullband speech,'' \emph{arXiv preprint arXiv:2008.04259}, 2020.

\bibitem{chen2020truth}
B.~Chen, H.~Wang, Y.~Wei, and R.~H. So, ``Truth-to-estimate ratio mask: A
  post-processing method for speech enhancement direct at low signal-to-noise
  ratios,'' in \emph{Proc. ICASSP}, 2020.

\bibitem{rahman2016adaptive}
S.~Rahman, M.~M. Rahman, M.~Abdullah-Al-Wadud, G.~D. Al-Quaderi, and
  M.~Shoyaib, ``An adaptive gamma correction for image enhancement,''
  \emph{EURASIP Journal on Image and Video Processing}, vol. 2016, no.~1, pp.
  1--13, 2016.

\bibitem{pavlovic2018sii}
C.~Pavlovic, ``Sii—speech intelligibility index standard: {ANSI} s3. 5
  1997,'' \emph{the Journal of the Acoustical Society of America}, vol. 143,
  no.~3, pp. 1906--1906, 2018.

\bibitem{poynton2012digital}
C.~Poynton, \emph{Digital video and {HD}: Algorithms and Interfaces}.\hskip 1em
  plus 0.5em minus 0.4em\relax Elsevier, 2012.

\bibitem{zhang2021low}
X.~Zhang, X.~Ren, X.~Zheng, L.~Chen, C.~Zhang, L.~Guo, and B.~Yu, ``Low-delay
  speech enhancement using perceptually motivated target and loss,'' in
  \emph{Proc. INTERSPEECH}, 2021.

\bibitem{singh1998speech}
L.~Singh and S.~Sridharan, ``Speech enhancement using critical band spectral
  subtraction,'' in \emph{Proc. ICSLP}, 1998.

\bibitem{liu2018bone}
H.-P. Liu, Y.~Tsao, and C.-S. Fuh, ``Bone-conducted speech enhancement using
  deep denoising autoencoder,'' \emph{Speech Communication}, vol. 104, pp.
  106--112, 2018.

\bibitem{fu2020boosting}
S.-W. Fu, C.-F. Liao, T.-A. Hsieh, K.-H. Hung, S.-S. Wang, C.~Yu, H.-C. Kuo,
  R.~Zezario, Y.-J. Li, S.-Y. Chuang \emph{et~al.}, ``Boosting objective scores
  of a speech enhancement model by metricgan post-processing,'' in \emph{Proc.
  APSIPA}, 2020.

\bibitem{ansi1997s3}
A.~ANSI, ``S3. 5-1997, methods for the calculation of the speech
  intelligibility index,'' \emph{New York: American National Standards
  Institute}, vol.~19, pp. 90--119, 1997.

\bibitem{zhao2018convolutional}
H.~Zhao, S.~Zarar, I.~Tashev, and C.-H. Lee, ``Convolutional-recurrent neural
  networks for speech enhancement,'' in \emph{Proc. ICASSP}, 2018.

\bibitem{fu2021metricgan+}
S.-W. Fu, C.~Yu, T.-A. Hsieh, P.~Plantinga, M.~Ravanelli, X.~Lu, and Y.~Tsao,
  ``Metricgan+: An improved version of metricgan for speech enhancement,''
  \emph{arXiv preprint arXiv:2104.03538}, 2021.

\bibitem{ravanelli2021speechbrain}
M.~Ravanelli, T.~Parcollet, P.~Plantinga, A.~Rouhe, S.~Cornell, L.~Lugosch,
  C.~Subakan, N.~Dawalatabad, A.~Heba, J.~Zhong \emph{et~al.}, ``Speechbrain: A
  general-purpose speech toolkit,'' \emph{arXiv preprint arXiv:2106.04624},
  2021.

\bibitem{dang2021dpt}
F.~Dang, H.~Chen, and P.~Zhang, ``Dpt-fsnet: Dual-path transformer based
  full-band and sub-band fusion network for speech enhancement,'' \emph{arXiv
  preprint arXiv:2104.13002}, 2021.

\bibitem{valentini2017noisy}
C.~Valentini-Botinhao \emph{et~al.}, ``Noisy speech database for training
  speech enhancement algorithms and {TTS} models,'' 2017.

\bibitem{pizer1987adaptive}
S.~M. Pizer, E.~P. Amburn, J.~D. Austin, R.~Cromartie, A.~Geselowitz, T.~Greer,
  B.~ter Haar~Romeny, J.~B. Zimmerman, and K.~Zuiderveld, ``Adaptive histogram
  equalization and its variations,'' \emph{Computer vision, graphics, and image
  processing}, vol.~39, no.~3, pp. 355--368, 1987.

\bibitem{qureshi1985adaptive}
S.~U. Qureshi, ``Adaptive equalization,'' \emph{Proceedings of the IEEE},
  vol.~73, no.~9, pp. 1349--1387, 1985.

\bibitem{koyama2020exploring}
Y.~Koyama, T.~Vuong, S.~Uhlich, and B.~Raj, ``Exploring the best loss function
  for {DNN-based} low-latency speech enhancement with temporal convolutional
  networks,'' \emph{arXiv preprint arXiv:2005.11611}, 2020.

\bibitem{defossez2020real}
A.~D{\'e}fossez, G.~Synnaeve, and Y.~Adi, ``Real time speech enhancement in the
  waveform domain,'' in \emph{Proc. INTERSPEECH}, 2020.

\end{thebibliography}

\end{document}